\documentclass[fleqn, 10pt]{wlscirep}
\usepackage[utf8]{inputenc}
\usepackage[T1]{fontenc}
\usepackage{graphicx}
\usepackage{verbatim}
\usepackage{fontawesome5}   
\definecolor{foldercolor}{RGB}{105,75,125}
\definecolor{jsoncolor}{RGB}{188,225,164}
\definecolor{datacolor}{RGB}{146,213,208}
\definecolor{eegcolor}{RGB}{209,101,82}
\usepackage{forest} 
\usepackage{makecell} 
\usepackage{xcolor}
\usepackage{dirtree}
\usepackage{multirow}
\usepackage{caption}
\usepackage{subcaption}

\title{A High-Density EEG Dataset for Stimulus-Driven Auditory Attention}

\author[1,2]{Ruofan Yan}
\author[1,*]{Na Lu}
\author[2]{Shu Peng}
\author[2]{Wenlong You}
\author[2]{Zhige Chen}
\author[1]{Yuxuan Yan}
\author[3]{Yan Liu}
\author[2]{Kay Chen Tan}
\author[2*]{Jibin Wu}
\affil[1]{Systems Engineering Institute, School of Automation Science and Engineering, Xi’an Jiaotong University, China}
\affil[2]{Department of Data Science and Artificial Intelligence, The Hong Kong Polytechnic University, Hong Kong SAR}
\affil[3]{Department of Computing, The Hong Kong Polytechnic University, Hong Kong SAR}

\affil[*]{jibin.wu@polyu.edu.hk, lvna2009@mail.xjtu.edu.cn}

\begin{abstract}
Elucidating how the human auditory system automatically prioritizes competing sound streams in complex environments is a central goal in both cognitive neuroscience and the development of auditory brain-computer interfaces (BCIs). To track these dynamic processes, electroencephalography (EEG)-based auditory attention decoding (AAD) has emerged as a crucial approach. However, most open-access EEG benchmarks focus on top-down, instructed target selection, falling short of capturing bottom-up, stimulus-driven orienting during instruction-free dichotic competition. To bridge this gap, we present the SAAD-EEG dataset, comprising 128-channel EEG recordings from 23 participants exposed to competing naturalistic sounds drawn from environmental, speech, and music domains. The experimental paradigm was deliberately designed without explicit target instructions, enabling the investigation of stimulus-driven prioritization based on acoustic and semantic salience. Alongside raw and preprocessed EEG data, the release includes trial-aligned orienting labels, reaction times, and subjective sound ratings. Comprehensive technical validation confirms the dataset's high neurophysiological quality, behavioral consistency, and robust decodability. This open-access resource opens new avenues for studying the neural mechanisms of bottom-up auditory attention and provides a critical foundation for developing attention-aware hearing technologies.

\end{abstract}
\begin{document}
    \flushbottom
    \maketitle
    %
    %
    \thispagestyle{empty}



    \section*{Background \& Summary}
    Recent progress in auditory BCIs and assistive listening devices has advocated  the use of neural decoding to determine which auditory stream is being prioritized by the user~\cite{belo2021eeg, zhang2025novel, zheng2025applications, wang2025auditory, han2019speaker,markovinovic2022design,hjortkjaer2025real}. EEG-based AAD has emerged as a promising, non-invasive, and temporally precise approach for tracking these attention-related neural dynamics~\cite{geirnaert2021electroencephalography, o2015attentional,kerlin2010attentional,li2024multi,wang2025research,das2020stimulus,straetmans2021neural}. However, most existing open-access EEG datasets have focused almost exclusively on goal-directed (top-down) selection, utilizing instructed paradigms wherein participants are explicitly directed to attend to a designated target audio stream~\cite{kaya2017modelling, mundanad2021eeg,das2016effect, deng2019impoverished,fuglsang2017noise,aroudi2019impact,zhang2023learnable, ceolini2020brain}. While these resources have facilitated benchmarking of top-down AAD algorithms~\cite{cai2025modeling, cai2024robust, pahuja2025, zeng2024attention, asjid2025single}, real-world communication frequently takes place in complex environments, where listeners are often interrupted by urgent or salient events, such as an infant crying or a fire alarm, that capture attention irrespective of current goals. Auditory BCIs incapable of tracking this stimulus-driven (bottom-up) orienting risk missing critical environmental cues. Despite the significance of this challenge, available EEG resources for stimulus-driven auditory attention remain scarce and are largely limited to detecting temporal salience within a single stream~\cite{tordini2015loud,bouvier2023revealing,kothinti2021auditory,kaya2014investigating,podwinska2019acoustic, straetmans2021neural,huang2017auditory,huang2019eeg}, rather than resolving instruction-free competition between multiple streams. Understanding how the brain manages these demands is crucial for developing comprehensive models of natural auditory attention, underscoring an urgent need for datasets and paradigms that reflect the true dynamics of unprompted, stimulus-driven auditory orienting in competing scenes.

    Here, we introduce the SAAD-EEG (Stimulus-Driven Auditory Attention EEG) dataset, which comprises 128-channel EEG recordings from 23 participants engaged in an instruction-free dichotic listening paradigm. The stimuli consist of naturalistic sound events categorized into eight subcategories and 24 cross-category pairing types, encompassing environmental sounds with both high and low ecological validity, speech, and music. In each trial, participants were presented with a dichotic pair of sounds without any explicit attentional cues. To minimize goal-directed selection and isolate genuine bottom-up orienting, they continuously watched a silent movie during sound presentation. After a brief post-stimulus delay, participants indicated which side was more attention-capturing. This design offers a unique opportunity to examine how competing audio streams draw attention based on their acoustic prominence, semantic meaning, and ecological relevance. 
    
\begin{figure}[ht]
        \centering
        \includegraphics[trim=1cm 18.5cm 0cm 1cm, scale=0.92]{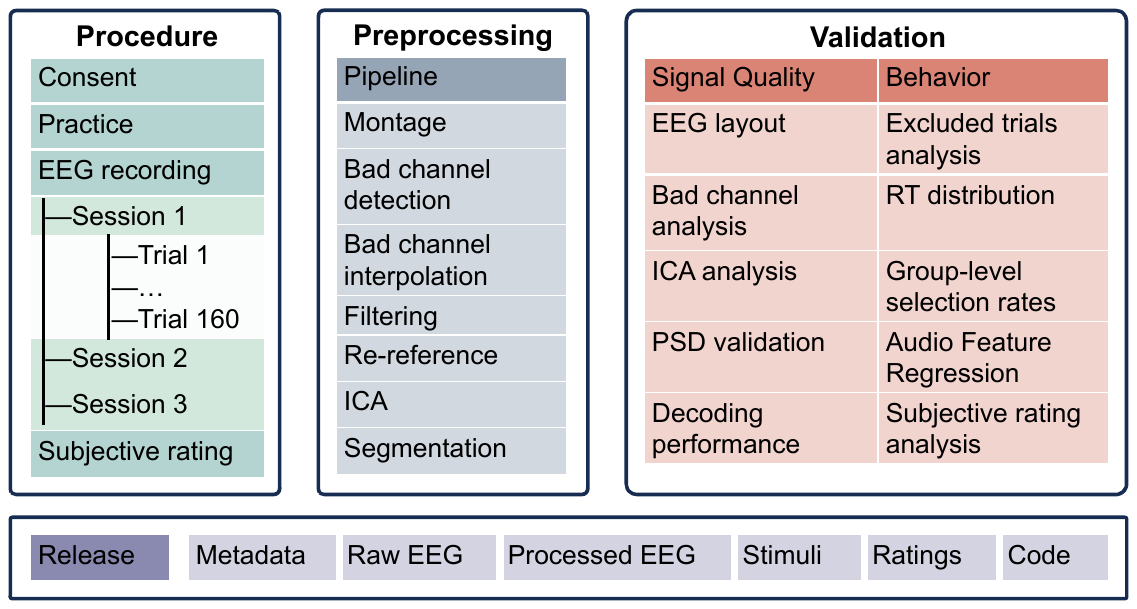}
        \caption{Schematic overview of the dataset generation and validation pipeline. The workflow comprises four stages: \textbf{Procedure}: After prescreening, participants provided informed consent, completed practice, and underwent formal EEG recording, followed by subjective ratings on representative sound clips. \textbf{Preprocessing}: EEG recordings were preprocessed according to standardized procedures. \textbf{Validation}: EEG data validation included signal quality assessment (layout verification, bad-channel analysis, ICA analysis), neurophysiological plausibility checks (PSD validation), and decoding performance evaluation. Behavioral validation included analyses such as excluded-trial summaries, reaction time distributions, and subjective rating assessments. \textbf{Release}: All data, including metadata, raw and processed EEG recordings, stimuli, subjective ratings, and processing scripts, were publicly released in a structured format.}
        \label{fig:scheme}
    \end{figure}

    Validation analyses confirm the dataset’s high quality, behavioral consistency, and robust decodability across multiple cross-validation schemes. The dataset also includes comprehensive behavioral annotations for the stimuli, including trial-level orienting-side labels, precise reaction times, and participant-level subjective ratings of each sound’s familiarity, urgency, and likability. These detailed annotations enable fine-grained analyses of how specific acoustic, semantic, and ecological properties influence neural processing and attentional capture in complex auditory environments. All data, including raw and preprocessed EEG files, stimulus manifests, and preprocessing scripts, are provided in standardized formats to ensure accessibility and reproducibility (see Figure~\ref{fig:scheme} for an overview of the dataset generation and validation pipeline).
    
    The SAAD-EEG dataset offers several advantages for advancing research in auditory neuroscience and neuroengineering. It facilitates investigation of the neural mechanisms underlying bottom-up auditory attention, illuminating how the brain automatically evaluates and prioritizes competing sound streams in the absence of explicit instruction. The inclusion of category-structured, naturalistic stimuli enables exploration of how ecological validity and semantic urgency influence stimulus-driven orienting. Ultimately, this reproducible and open-access resource provides a foundation for developing next-generation, attention-aware hearing devices that can seamlessly adapt to the complexities of real-world acoustic environments.

\section*{Methods}

\subsection*{Participants}

    The dataset comprises EEG recordings from 23 adult participants (14 males and 9 females), with a mean age of 27.0 $\pm$ 2.7 years. All participants were native Chinese speakers, fluent in English, and enrolled as students at The Hong Kong Polytechnic University. 22 participants were right-handed and one participant was left-handed. All participants had normal hearing and reported no history of neurological disorders. The study was approved by the Institutional Review Board of The Hong Kong Polytechnic University (Reference No. HSEARS20230628001). Before the experiment, participants received a detailed explanation of the experimental procedure, potential risks, and task-related requirements. Each participant provided written informed consent and agreed to the voluntary collection and public release of their data. They were compensated for their time and contribution.


\begin{table}[t]
  \centering
  \caption{Hierarchical taxonomy of stimulus subcategories.}
  \label{tab:subcategories}
  \begin{tabular}{|p{4.6cm}|p{5.4cm}|c|}
    \hline
    \textbf{Main Categories} & \textbf{Subcategories} & \textbf{Code} \\
    \hline
    \multirow{2}{=}{High-ecological environmental sounds}
      & Baby cry, infant cry & A1 \\
    \cline{2-3}
      & Telephone bell ringing & A2 \\
    \hline
    \multirow{2}{=}{Low-ecological environmental sounds}
      & Computer keyboard & B1 \\
    \cline{2-3}
      & Helicopter & B2 \\
    \hline
    \multirow{2}{=}{Speech}
      & Male speech, man speaking & C1 \\
    \cline{2-3}
      & Female speech, woman speaking & C2 \\
    \hline
    \multirow{2}{=}{Music}
      & Bass drum & D1 \\
    \cline{2-3}
      & Sad music & D2 \\
    \hline
  \end{tabular}
\end{table}

\begin{table}[t]
  \centering
  \caption{Twenty-four subcategory-level dichotic pairing types derived from cross-category combinations. See Table~\ref{tab:subcategories} for code definitions.}
  \label{tab:pairing_types}
  \begin{tabular}{|l|p{9.5cm}|}
    \hline
    \textbf{Main-category pair} & \textbf{Subcategory pairing type codes} \\
    \hline
    A--B & A1B1, A1B2, A2B1, A2B2 \\
    \hline
    A--C & A1C1, A1C2, A2C1, A2C2 \\
    \hline
    A--D & A1D1, A1D2, A2D1, A2D2 \\
    \hline
    B--C & B1C1, B1C2, B2C1, B2C2 \\
    \hline
    B--D & B1D1, B1D2, B2D1, B2D2 \\
    \hline
    C--D & C1D1, C1D2, C2D1, C2D2 \\
    \hline
  \end{tabular}
\end{table}
\subsection*{Stimuli}
Stimuli were constructed as dichotic audio pairs, simultaneously presented to the left and right ears. Source clips were selected from AudioSet \cite{gemmeke2017audio}, a manually annotated corpus of audio events sourced from YouTube, to ensure broad coverage of naturalistic, real-world sounds. The stimulus pool was organized according to a hierarchical taxonomy comprising four main categories: high-ecological (HE) and low-ecological (LE) environmental sounds, speech, and music, each subdivided into two subcategories. Ecological validity was defined in terms of everyday occurrence and attentional relevance, facilitating comparisons across distinct sound categories. Complete subcategory definitions and codes are provided in Table~\ref{tab:subcategories}.

Clips were manually screened, and 2-s excerpts were carefully selected to maximize label consistency while preserving perceptual clarity. The amplitude of each clip was normalized based on the root mean square (RMS) amplitude calculated from the loudest 1\% of samples, reducing potential loudness-related confounds across stimulus pairs. Pair construction was restricted to cross-category combinations, with within-category and self-pairs excluded. Under these constraints, 24 valid subcategory-level pairing types were generated, as enumerated in Table~\ref{tab:pairing_types}. Each pairing type was repeated 10 times and expanded with mirrored left–right versions to balance ear assignment and limit side confounds, resulting in a total of 480 dichotic stimulus pairs.

\begin{figure}[ht]
    \centering
    \begin{subfigure}{\linewidth}
        \includegraphics[trim=0.1cm 16.5cm 0cm 1.5cm, scale=0.85]{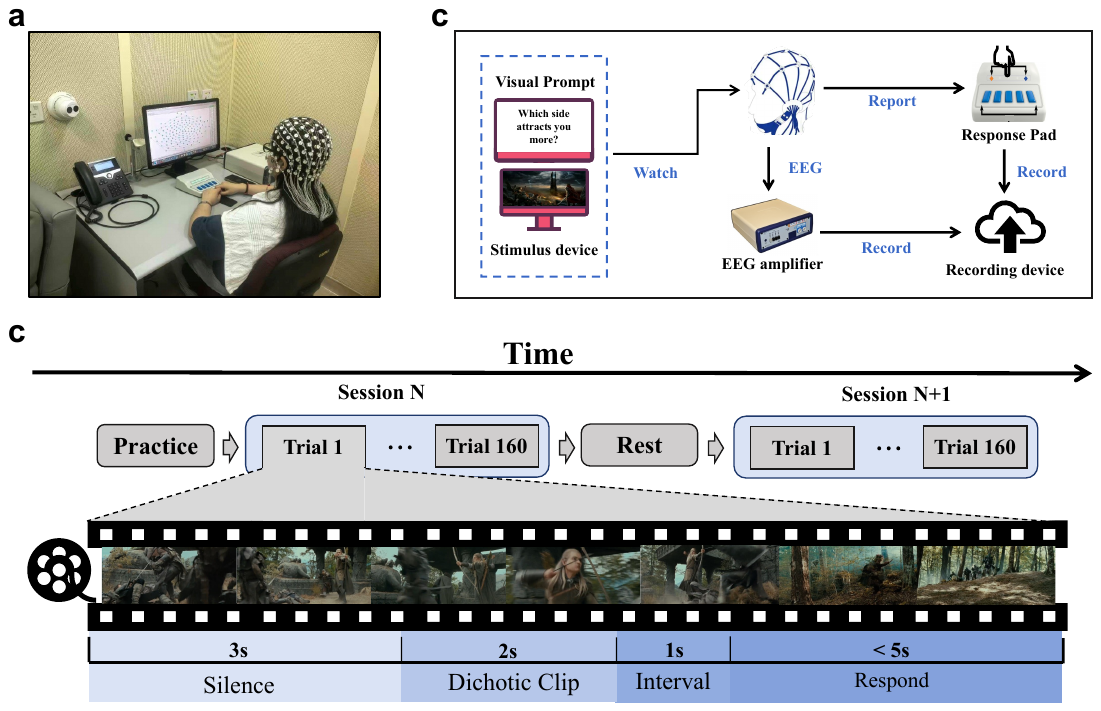}
        \phantomsubcaption\label{fig:procedure2a}
        \phantomsubcaption\label{fig:procedure2b}
        \phantomsubcaption\label{fig:procedure2c}
    \end{subfigure}
    \caption{
        (a) The EEG experimental setup.
        (b) Hardware configuration and data flow.
        (c) Overview of the experimental procedure.
    }
    \label{fig:procedure}
\end{figure}
    \subsection*{Experimental Procedure}
     All recordings were conducted in a sound-attenuated and electromagnetically shielded room. As illustrated in Figures~\ref{fig:procedure2a} and \ref{fig:procedure2b}, participants wore an EEG cap and wired headphones, viewed a monitor positioned approximately 50 cm in front, and responded using a response pad. Saline was applied to ensure stable conductivity prior to recording. A five-trial practice block preceded the main acquisition phase. During the main phase, the 480 dichotic stimulus pairs were randomly divided into three sessions (160 trials per session). To minimize fatigue and maintain data quality, adequate rest was provided between sessions, during which saline was replenished as needed to preserve electrode contact. 
     
     At the beginning of each session, a central fixation cross was presented to prompt participants to maintain attention. Participants initiated the run by pressing the response pad. As shown in Figure~\ref{fig:procedure2c}, each trial began with a 3-second silent period, ensuring that each dichotic stimulus pair started from silence. The two sound clips were then simultaneously presented to the left and right ears for 2 seconds without ear-specific attentional instruction. A 1-second post-stimulus delay preceded response collection to reduce potential contamination of stimulus-related EEG activity by motor responses associated with button presses \cite{luck2005ten}. After this delay, a response prompt instructed participants to report which side was more attention-capturing by pressing the corresponding left or right key. Participants were allowed up to five seconds to respond, and trials exceeding this response window were excluded from subsequent analysis. Trial labels therefore reflected participants’ post-stimulus self-reports of the more attention-capturing side, rather than externally instructed or objectively verified attended ears. Throughout each session, a silent American feature film was displayed. Participants were instructed to follow the movie and not to deliberately monitor the dichotic sounds. This design was intended to minimize explicit goal-directed selection and capture stimulus-driven auditory orienting during dichotic competition. 
        
    \subsection*{Data Acquisition}
    EEG signals were recorded using a 128-channel HydroCel GSN system (Electrical Geodesics, Inc.) at a sampling rate of 250 Hz. Electrode positions followed the international 10–10 system, and impedance was maintained below 10 k$\Omega$ before recording. During acquisition, signals were referenced online to the vertex (Cz), with the COM sensor (posterior to Cz) used as ground. Stimulus presentation, event-marker delivery, and response logging were implemented in E-Prime 2.0 (Psychology Software Tools, Pittsburgh, PA). Behavioral responses were collected via the EGI Chronos response pad, which recorded left/right choices and reaction time (RT) with millisecond precision. Through the E-Prime–Chronos interface, trial events, behavioral responses, and EEG recordings were temporally aligned at the trial level. In addition, participants completed post-experiment subjective ratings using an online questionnaire. For each subcategory, one representative audio clip was presented, and participants rated familiarity, liking, and urgency on a 0–4 scale. Each item was rated after listening to its corresponding clip.
    At the end of the experiment, raw EEG data together with event markers were saved in \texttt{.mff} format for offline analysis, and behavioral logs were exported as \texttt{.txt} files.
\begin{figure}[!thbp]
  \centering
  \includegraphics[width=0.75\linewidth]{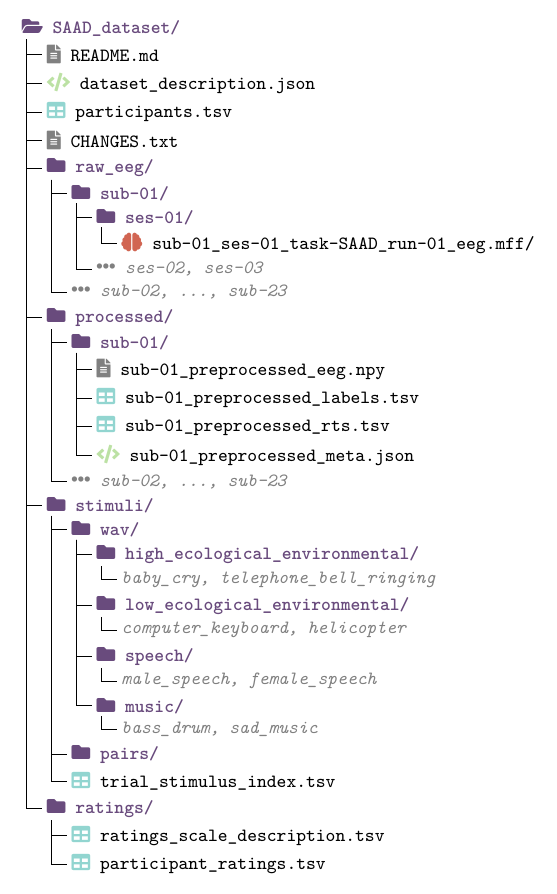}

\caption{Directory layout of the SAAD-EEG dataset as released on Zenodo. Icons indicate item types: \textcolor{foldercolor}{\faIcon{folder}}~directories, \textcolor{eegcolor}{\faIcon{brain}}~raw EEG recordings (\texttt{.mff}), \textcolor{jsoncolor}{\faIcon{code}}~JSON metadata, \textcolor{datacolor}{\faIcon{table}}~tabular files (\texttt{.tsv}), and \textcolor{gray}{\faIcon{file-alt}}~other release files (\texttt{.md}/\texttt{.npy}). Only \texttt{sub-01} is shown in full. Ellipses (...) indicate repeated structures across participants and sessions. Under each main audio category in \texttt{stimuli/wav/}, grey italic text lists the two subcategory folder names.}
\label{fig:dataset_tree}
\end{figure}

    \subsection*{Data preprocessing}
     EEG preprocessing was performed offline in Python (v3.12.7) using MNE-Python (v1.10.2 \cite{gramfort2013meg}), PyPREP (v0.4.3 \cite{bigdely2015prep}) and mne-icalabel \cite{pion2019iclabel}. Raw recordings were configured with the GSN-HydroCel-128 montage, with VREF treated as a miscellaneous channel. Noisy channels were automatically identified using PyPREP  based on local correlation criteria and then interpolated using spherical spline interpolation. The data were subsequently re-referenced to the common average reference. A 0.1-40 Hz FIR band-pass filter was applied to remove DC offset and suppress slow baseline drifts, while retaining frequency components relevant to auditory attention. A 50 Hz notch filter was then applied to attenuate line noise. For artifact reduction, independent component analysis (ICA) with the Infomax algorithm \cite{bell1995information} was used to decompose the EEG into independent components. The ICLabel classifier (mne-icalabel) was then applied to assign each component to categories such as brain, eye, and muscle; non-brain components were removed, and the cleaned EEG was reconstructed from the remaining components. Continuous EEG was segmented into 6\,s trial epochs locked to trial onset. For analysis and deposited processed data, time is expressed relative to dichotic-stimulus onset ($t=0$\,s): a 3\,s pre-stimulus silent baseline ($-3$ to $0$\,s), a 2\,s dichotic-stimulation interval ($0$ to $2$\,s), and a 1\,s post-stimulus silent interval ($+2$ to $+3$\,s) before the orienting response. Baseline correction for validation analyses used the $-3$ to $0$\,s window; primary decoding and spectral summaries used the $0$ to $2$\,s stimulation window. Released arrays store the same 6\,s segments with stimulus onset at sample index 750 (3\,s $\times$ 250\,Hz); see Data Records for array layout.

\section*{Data Records}
The dataset is publicly available on the Zenodo platform \cite{yan2026saadeeg}. The release follows a hierarchical organization designed for transparent reuse. As shown in Figure~\ref{fig:dataset_tree}, the dataset includes root-level metadata files and four main subdirectories: raw\_eeg/ (raw EEG recordings), processed/ (preprocessed EEG data), stimuli/ (stimulus audio files), and ratings/ (participant-level subjective ratings). For clarity, the figure shows the full tree for \texttt{sub-01} only; all other participants and the remaining sessions follow the same file naming and folder structure.

\subsection*{Root-level metadata}
The \texttt{README.md} file describes the Zenodo folder layout and refers readers to preprocessing code on GitHub (Code availability). The \texttt{dataset\_description.json} file records technical parameters of the dataset, including sampling rate, channel count, trial count, and epoch timing. Participant ID, sex, age, and handedness are listed in \texttt{participants.tsv}. The data contains no personally identifiable information beyond these demographics. The \texttt{CHANGES.txt} file logs changes between dataset releases.

\subsection*{Raw EEG folder}
The \texttt{raw\_eeg/} directory stores continuous session-level EEG recordings in EGI \texttt{.mff} format, organized by participant (\texttt{sub-XX/}) and session (\texttt{ses-01/}, \texttt{ses-02/}, \texttt{ses-03/}). Each session folder contains one recording, \texttt{sub-XX\_\allowbreak ses-YY\_\allowbreak task-SAAD\_\allowbreak run-01\_\allowbreak eeg.mff/}, as acquired during data collection.
\subsection*{Processed folder}
The \texttt{processed/} directory contains trial-epoched EEG and behavioral data per participant. The three sessions are merged: each \texttt{sub-XX/} folder contains 480 trials in four files. The file \texttt{sub-XX\_preprocessed\_eeg.npy} stores epoched EEG as a NumPy array of shape (480, 128, 1500) at 250 Hz, corresponding to 6 s per trial spanning the pre-stimulus, dichotic-stimulation, and post-stimulus intervals of each trial. The files \texttt{sub-XX\_preprocessed\_labels.tsv} and \texttt{sub-XX\_preprocessed\_rts.tsv} provide trial-wise orienting-side labels (\texttt{Label}: 0 = left, 1 = right) and reaction times (\texttt{RT\_ms}). The file \texttt{sub-XX\_preprocessed\_meta.json} records channel names, preprocessing information, and epoch timing.

\subsection*{Stimuli folder}

The \texttt{stimuli/} directory contains the auditory materials used in the experiment. The \texttt{wav/} subfolder contains single-ear reference clips organized by four main categories, each with two subcategory folders. The \texttt{pairs/} subfolder holds 480 dichotic pair waveforms (\texttt{.wav}), one file per experimental pair. The file \texttt{trial\_stimulus\_index.tsv} lists the playback order for all sessions (480 rows: 160 per session) with columns \texttt{session}, \texttt{stimulus\_index}, \texttt{pair\_id}, and \texttt{pair\_relative\_path}, linking each trial to the corresponding file in \texttt{pairs/}.

\subsection*{Ratings folder}

The \texttt{ratings/} directory contains two tab-separated files. \texttt{ratings\_scale\_description.tsv} defines the post-experiment rating scales (familiarity, urgency, and liking; 0--4) with English questions and anchor labels. \texttt{participant\_ratings.tsv} contains the ratings for each participant across eight sound subcategories.

\begin{figure}[th]
    \centering
    \begin{subfigure}{\linewidth}
        \includegraphics[trim=0cm 14.5cm 0.5cm 1.5cm, scale=0.85]{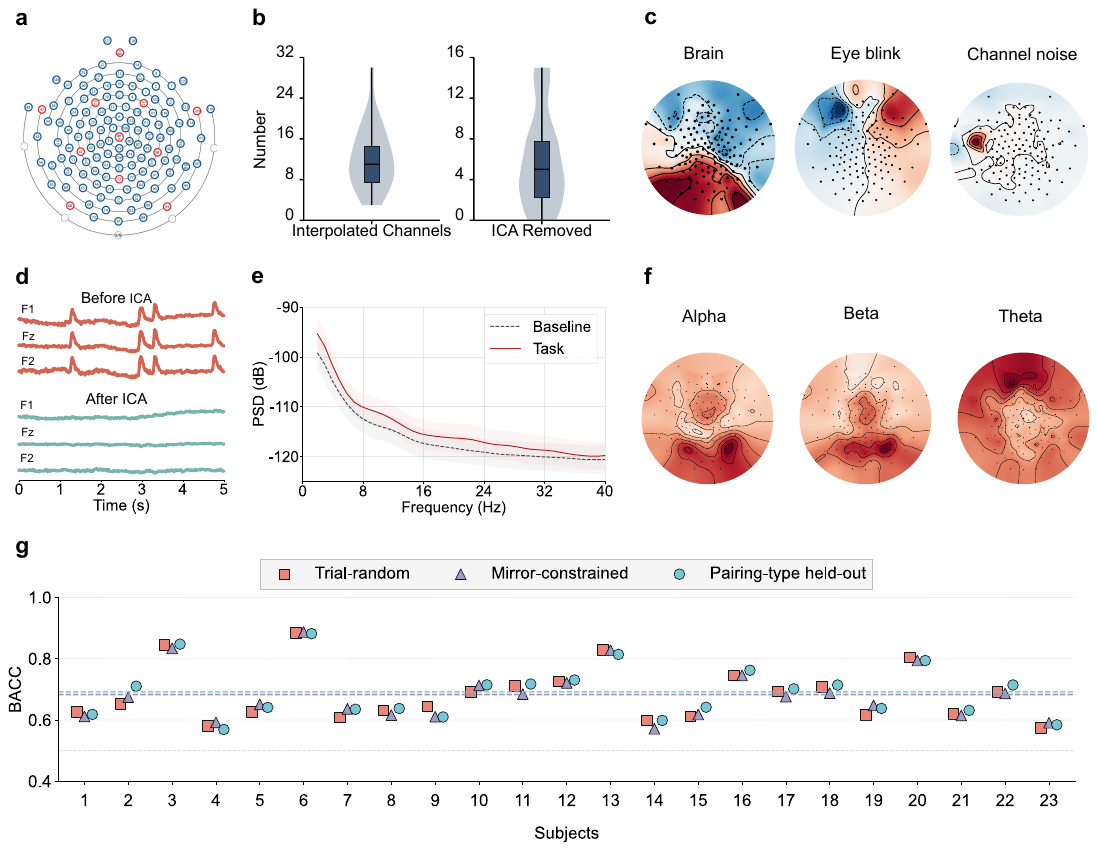}
        \phantomsubcaption\label{fig:signal_a}
        \phantomsubcaption\label{fig:signal_b}
        \phantomsubcaption\label{fig:signal_c}
        \phantomsubcaption\label{fig:signal_d}
        \phantomsubcaption\label{fig:signal_e}
        \phantomsubcaption\label{fig:signal_f}
        \phantomsubcaption\label{fig:signal_g}
    \end{subfigure}
    \caption{
Signal quality assessment results for the SAAD-EEG dataset.
(a) Sensor layout of the 128-channel HydroCel EEG system.
(b) Number of interpolated channels and excluded ICA components for each participant.
(c) Representative independent components identified by ICA and labeled with ICLabel.
(d) EEG signals at representative electrodes F1, Fz, and F2 before and after artifact removal.
(e) Group mean PSD (dB) over 1--40 Hz for the pre-stimulus baseline (-3 to 0 s; dashed grey) and dichotic-stimulus (0 to 2 s; solid red) windows. Shaded regions indicate the standard error of the mean (SEM) across participants.
(f) Scalp topographies of the stimulus-minus-baseline log-power difference (dB) for theta, alpha, and beta bands, averaged across participants.
(g) Robustness of EEGNet decoding for each participant across three cross-validation schemes: trial-random, mirror-constrained, and category-held-out splits. The dashed horizontal line indicates chance level.
}
    \label{fig:signal}
\end{figure}

    \section*{Technical Validation}
    To assess the reliability and quality of the dataset, we performed a series of validation analyses covering EEG signal quality, neurophysiological plausibility, decoding performance and robustness across partitioning schemes, and behavioral consistency.

    \subsection*{Signal Quality Assessment}
We first assessed EEG quality after preprocessing to support reproducible reuse of the dataset. Figure~\ref{fig:signal_a} presents the 128-channel HydroCel sensor layout used for data acquisition, providing dense scalp coverage for spatially resolved EEG analyses. The number of interpolated channels for each participant is summarized in Figure~\ref{fig:signal_b} (left). Across participants, interpolation was required for 8–15 channels (median = 11). One participant required 30 interpolated channels. This outlier may reflect individual differences in electrode contact quality and cap fitting during acquisition. Excluded ICA-component counts per participant are summarized in Figure~\ref{fig:signal_b} (right). Across participants, 3–9 ICA components were removed (median = 5). These interpolation and ICA-rejection counts fall within ranges commonly observed in high-density saline-based EEG recordings, supporting the technical usability of the deposited EEG. Figure~\ref{fig:signal_c} shows representative independent components identified by ICA and labeled with ICLabel, illustrating typical retained brain activity alongside common rejected artifact components such as eye blinks and channel noise. Figure~\ref{fig:signal_d} compares EEG signals at representative frontocentral electrodes (F1, Fz, and F2) before and after ICA removal, demonstrating reduction of blink-related artifacts while preserving the overall temporal characteristics of the EEG signals.

\subsection*{Neurophysiological Plausibility Check}
To assess neurophysiological plausibility, we compared group-level spectral profiles and scalp topographies between the baseline window (-3 to 0 s) and the dichotic-stimulation window (0 to 2 s). Group mean power spectral density (PSD) across participants is shown in Figure~\ref{fig:signal_e}. Relative to baseline, the stimulation window showed increased spectral power across 0–40 Hz, with larger changes observed in lower-frequency bands. Theta (4–8 Hz) and alpha (8–13 Hz) exhibited the strongest increases, whereas beta (13–30 Hz) showed a smaller but consistent increase. These results indicate stimulus-related spectral modulation relative to the pre-stimulus baseline. To further map the spatial distribution of spectral changes, we computed scalp topographies of the stimulus-minus-baseline log-power difference for alpha, beta and theta bands, as shown in Figure~\ref{fig:signal_f}. The topographic maps showed distinct spatial distributions across frequency bands. Alpha and beta increases were most prominent over parieto-occipital regions, broadly consistent with previous reports of posterior task-related modulation\cite{deng2020topographic}. Theta increases were most prominent over anterior frontal regions, broadly consistent with previously reported frontal theta involvement in cognitive control and engagement processes \cite{cavanagh2014frontal}. Together, these patterns support the neurophysiological plausibility of the released EEG dataset.

    \subsection*{Baseline Decoding Results}
    To evaluate the decodability of the released EEG dataset, we employed EEGNet \cite{lawhern2018eegnet}, a compact deep learning model designed for EEG-based decoding tasks, as a baseline classifier. Using depthwise separable convolutions, EEGNet learns spatial and temporal features efficiently with fewer parameters. This model has been successfully applied to auditory attention decoding tasks\cite{mahjoory2024convolutional,wang2025research}.

    To ensure a fair and reproducible model evaluation, we adopted a within-subject, five-fold cross-validation strategy. For each participant, trials were randomly shuffled and partitioned into five approximately equal-sized folds. Each fold was used once as the test set, while the remaining four folds were used for training. The same partitioning procedure was applied to all participants, and fold assignments were generated using a fixed random seed to enable reproducibility. Because the labels represent participants’ self-reported orienting choices, class distributions may be imbalanced and vary across individuals. Therefore, balanced accuracy (BACC) and standard deviation (SD) were reported to provide equal weighting across classes. For each participant, BACC was averaged across the five test folds.

    As shown in Table \ref{tab:bacc_split}, EEGNet achieved above-chance trial-wise prediction of the response labels under the random split cross-validation scheme. At the group level, balanced accuracy averaged 68.38\% (SD = 9.5\% across participants), indicating reliable but moderate decodability on this stimulus-driven listening task. Individual performance varied across participants, with mean BACC values ranging from approximately 57\% to 88\%. However, within-participant variability across cross-validation folds was relatively small for most individuals, indicating stable evaluation across different test splits. Overall, these results demonstrate the presence of discriminative neural information in the released EEG data while highlighting inter-individual variability commonly observed in EEG decoding studies.

    \begin{table}[t]
  \centering
  \small
  \caption{Each cell reports per-participant BACC and fold-wise SD.}
  \label{tab:bacc_split}
  \setlength{\tabcolsep}{4pt}
  \renewcommand{\arraystretch}{1.15}
  \begin{tabular}{|c|c|c|c|c|c|c|c|c|c|c|c|}
    \hline
    S1 & S2 & S3 & S4 & S5 & S6 & S7 & S8 & S9 & S10 & S11 & S12 \\
    \hline
    \makecell{0.6262\\$\pm$0.0296} & \makecell{0.6539\\$\pm$0.0376} & \makecell{0.8442\\$\pm$0.0291} & \makecell{0.5820\\$\pm$0.0570} & \makecell{0.6271\\$\pm$0.0320} & \makecell{0.8837\\$\pm$0.0251} & \makecell{0.6088\\$\pm$0.0219} & \makecell{0.6317\\$\pm$0.0249} & \makecell{0.6443\\$\pm$0.0226} & \makecell{0.6924\\$\pm$0.0524} & \makecell{0.7121\\$\pm$0.0254} & \makecell{0.7261\\$\pm$0.0287} \\
    \hline
    S13 & S14 & S15 & S16 & S17 & S18 & S19 & S20 & S21 & S22 & S23 & \textbf{Avg} \\
    \hline
    \makecell{0.8297\\$\pm$0.0239} & \makecell{0.5988\\$\pm$0.0191} & \makecell{0.6122\\$\pm$0.0449} & \makecell{0.7458\\$\pm$0.0356} & \makecell{0.6936\\$\pm$0.0305} & \makecell{0.7089\\$\pm$0.0380} & \makecell{0.6172\\$\pm$0.0342} & \makecell{0.8044\\$\pm$0.0240} & \makecell{0.6194\\$\pm$0.0342} & \makecell{0.6916\\$\pm$0.0433} & \makecell{0.5733\\$\pm$0.0141} & \makecell{\textbf{0.6838}\\\textbf{$\pm$0.0949}} \\
    \hline
  \end{tabular}
\end{table}

    \subsection*{Robustness across partitioning strategies}
    Beyond the trial-level random split, we designed two additional cross-validation schemes to assess whether decoding performance was driven by trial-level neural information rather than shortcut learning from ear-of-presentation cues or memorized stimulus pairings. All analyses were conducted within-subject, with only the rule for grouping trials into folds differing across schemes.

    \textbf{Mirror-constrained split.} Random splitting may introduce shared acoustic information across training and testing folds, allowing decoders to exploit ear-of-presentation or stimulus-specific features. To reduce this potential leakage, mirrored trials from the same pair were assigned to the same fold, ensuring that the model never encountered the opposite ear assignment of a given stimulus pair during training. Five-fold cross-validation was performed under this grouping strategy.

\textbf{Pairing-type held-out split.}
Models may learn specific subcategory pairing patterns rather than generalize to unseen stimulus combinations. We therefore adopted a six-fold cross-validation scheme based on the 24 subcategory-level pairing types defined in the stimulus taxonomy. In each split, models were trained on 20 pairing types and evaluated on 4 held-out pairing types. This scheme evaluates generalization to unseen stimulus pairings rather than leave-one-category-out generalization to entirely novel sound categories.

For each scheme, we trained EEGNet with the same preprocessing and hyperparameters, summarized per-participant BACC, and then reported group-level means across participants.
As shown in Figure~\ref{fig:signal_g}, the three cross-validation schemes yielded comparable decoding performance: random split (68.4\%), mirror-constrained split (68.2\%), and pairing-type held-out split (68.3\%).
The small performance differences across schemes suggest that decoding was not primarily driven by ear-of-presentation cues or memorization of specific stimulus pairings. These results indicate that the released EEG data contain trial-level neural information related to participants’ self-reported stimulus-driven orienting outcomes.
    \begin{figure}[h]
    \centering
    \begin{subfigure}{\linewidth}
        
    \includegraphics[trim=0.3cm 20cm 1.5cm 1cm, scale=0.9]{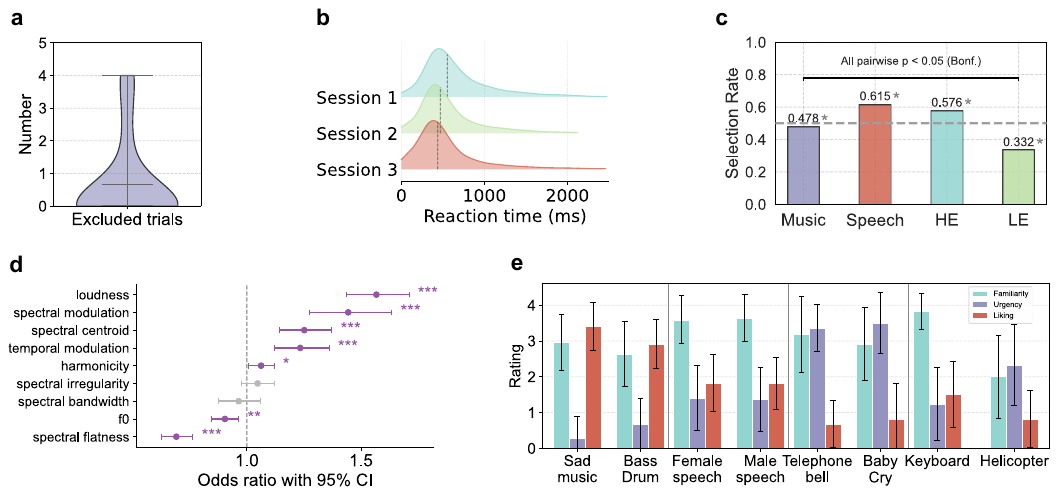}
    \phantomsubcaption\label{fig:behaviour_a}
    \phantomsubcaption\label{fig:behaviour_b}
    \phantomsubcaption\label{fig:behaviour_c}
    \phantomsubcaption\label{fig:behaviour_d}
    \phantomsubcaption\label{fig:behaviour_e}
    \end{subfigure}
    \caption{
Behavioral analysis results for the SAAD-EEG dataset.
(a) Number of excluded trials for each participant. (b) RT distributions for three sessions across all participants. (c) Group-level selection rates for music, speech, high-ecology, and low-ecology sounds in the dichotic listening stimulus-driven attention task. Each bar shows the ratio of trials in which a category was selected to trials in which it was available as the left- or right-ear stimulus. The dashed line at 0.5 indicates chance performance (no systematic bias). Gray asterisks mark categories whose selection rates differ significantly from 0.5; the horizontal bracket and label indicate that all pairwise comparisons between categories are significant. Bonferroni correction ($\alpha= 0.05$) was applied separately to each test family. (d) Pair-level logistic regression (binomial GLM, logit link). Odds ratios (OR) and 95\% confidence intervals for left--right audio-feature differences and group-level selection on one side of each stimulus pair. The dashed vertical line indicates OR $= 1$; $^{*}p < 0.05$, $^{**}p < 0.01$, $^{***}p < 0.001$. (e) Mean subjective ratings across participants (familiarity/liking/urgency; 0--4 scale)  of representative clips for each subcategory.
}
    \label{fig:behavior}
\end{figure}

\subsection*{Behavioral Data Validation}
This section presents technical validation of the behavioral data through response quality, category-level response patterns, and complementary acoustic and subjective characterizations. Figure~\ref{fig:behaviour_a} summarizes trial exclusions under the predefined RT criterion (>5,000 ms). In total, only 15 trials were removed, representing 0.14\% of all trials, indicating high completeness of the behavioral responses. Figure~\ref{fig:behaviour_b} presents RT distributions across the three sessions. Each session showed a typical positive skew, with about 60\% of trials between 300–800 ms and a right tail of slower responses. Median RT decreased progressively from Session 1 (554 ms) to Session 2 (469 ms) and Session 3 (440 ms), indicating increasing task familiarity and minimal fatigue effects. 

To examine whether behavioral responses exhibited non-random selection patterns across the predefined sound categories, we computed selection rates (SR) for the four main sound categories (music, speech, HE, and LE). For each category, we defined the selection rate as: $\text{SR}_c = \frac{n_c^{\text{selected}}}{N_c^{\text{presented}}}$,
where \(c\) denotes the sound category, \(n_c^{\text{selected}}\) and \(N_c^{\text{presented}}\) denote the numbers of selected and presented trials for category \(c\), respectively. Under unbiased selection, the expected selection rate is 0.5.
Figure~\ref{fig:behaviour_c} shows that all four categories deviated significantly from the expected value of 0.5 after Bonferroni correction. Speech (SR = 0.61) and HE (SR = 0.58) were selected more frequently than expected, whereas music (SR = 0.48) and LE (SR = 0.33) were selected less frequently. This trend is broadly consistent with the greater semantic salience and ecological relevance of speech and HE sounds, whereas music and LE environmental sounds may have been perceived as more background-like. All pairwise comparisons between categories were significant, indicating distinct category-level behavioral response patterns.

We further examined whether behavioral choices were systematically associated with measurable acoustic differences between competing sounds. Pair-level binomial GLMs (logit) were fitted using the left–right differences (\(\Delta\)) for nine perceptual features defined in a previous study\cite{huang2017auditory}. Each feature was \(z\)-scored across pairs. As shown in Figure~\ref{fig:behaviour_d}, loudness, spectral modulation, spectral centroid, and temporal modulation were positively associated with behavioral choices (all \(p < 0.001\)), whereas spectral flatness and \(f_0\) showed negative associations (\(p < 0.001\) and \(p < 0.01\), respectively). Harmonicity showed a weak positive association (\(p < 0.05\)), while the remaining features were not significant.

Figure~\ref{fig:behaviour_e} presents the post-experiment subjective ratings of familiarity, perceived urgency, and liking across the eight stimulus subcategories. Within each main category, the two subcategories showed broadly similar subjective rating profiles. Across categories, music received the highest liking ratings and the lowest urgency ratings, whereas HE sounds showed the opposite pattern. Speech was relatively familiar, and LE sounds generally received lower liking ratings. These subjective profiles were broadly consistent with the expected perceptual characteristics of the predefined stimulus categories.

\subsection*{Usage Notes}

The SAAD-EEG dataset provides a structured and reproducible resource for studies of stimulus-driven auditory attention, combining high-density EEG recordings with rich behavioral, stimulus, and subjective annotations. Researchers interested in customized preprocessing or validating alternative preprocessing strategies may use the raw \texttt{.mff} recordings, whereas the released preprocessed EEG epochs are intended for direct statistical analysis and machine learning. The released epochs preserve the complete trial structure, enabling flexible analyses of pre-stimulus, stimulus-evoked, and post-stimulus neural activity. We recommend MNE-Python (validated with version 1.10.2) to ensure compatibility with the released preprocessing pipeline.

Unlike conventional auditory attention datasets based on explicit listening instructions, the released labels represent participants' self-reported orienting outcomes during instruction-free dichotic competition. They should therefore be interpreted as behavioral measures of stimulus-driven attentional orienting rather than objective measures of attentional state.

Beyond serving as a reference benchmark for stimulus-driven auditory attention decoding, the released dataset supports several additional research directions. First, the high-density EEG recordings collected under an instruction-free paradigm provide a valuable resource for investigating the neural mechanisms of stimulus-driven auditory attention, complementing the extensive literature based on instructed attention paradigms. Second, the combination of behavioral annotations, stimulus metadata, and subjective ratings enables investigations of how acoustic, semantic, and ecological sound properties influence attentional orienting and its neural correlates. Finally, the dataset supports the development and evaluation of multimodal computational models of stimulus-driven auditory attention.

\section*{Data Availability}
The SAAD-EEG dataset is publicly accessible on the Zenodo platform at \url{https://zenodo.org/records/20557225} (DOI: 10.5281/zenodo.20557225.).

\section*{Code Availability}
The preprocessing and analysis scripts described in this paper are publicly available on GitHub (\url{https://github.com/yanruofan628/saad-preprocessing}).

\bibliography{sample}

\section*{Acknowledgements}

This work was partially supported by the Research Grants Council of the Hong Kong SAR (Grant No. C5052-23G, PolyU15215623, PolyU15229824, PolyU25216423, PolyU15217424, and SRFS2526-5S04), the National Natural Science Foundation of China (Grant No. 62306259), and The Hong Kong Polytechnic University (P0056503, P0058445, and P0060651). This work was also supported by the National Natural Science Foundation of China (Grant No. 62476213) and the Natural Science Basic Research Program of Shaanxi Province (Grant No. 2024JC-YBMS-486).

\section*{Author contributions}

R.Y. and J.W. conceived the dataset and designed the experiment. R.Y., S.P., W.Y. collected the EEG and behavioral data. R.Y., W.Y., Z.C., Y.Y. performed preprocessing, dataset curation, and technical validation analyses. R.Y. drafted the manuscript. J.W., N.L., W.Y., S.P. contributed to results discussion and manuscript writing. J.W., K.T., N.L., Y.L. supervised the whole project.

\section*{Competing interests}
The authors declare no competing interests. 

\end{document}